\long\def\symbolfootnote[#1]#2{\begingroup%
\def\thefootnote{\fnsymbol{footnote}}\footnote[#1]{#2}\endgroup} 
\def\aj{AJ}
\def\apj{ApJ}
\def\apjl{ApJ}
\def\apjs{ApJS}
\def\aap{A\&A}
\def\aaps{A\&AS}
\def\mnras{MNRAS}
\def\nat{Nature}

\newcommand{\chifit}{\chi_T}
\newcommand{\dr}{\rm{d}}
\newcommand{\edot}{\dot{E}}
\newcommand{\edotsev}{\dot{E}_{\rm SEV}}
\newcommand{\gyr}{{\rm Gyr}}
\newcommand{\henon}{H{\'e}non}

\newcommand{\mmax}{m_{\rm max}}
\newcommand{\mmin}{m_{\rm min}}
\newcommand{\mmean}{\bar{m}}
\newcommand{\mmeann}{\bar{m}_0}
\newcommand{\msun}{\rm{M}_{\odot}}
\newcommand{\msunpc}{{\rm M}_{\odot}\,{\rm pc}^{-3}}

\newcommand{\myr}{{\rm Myr}}

\newcommand{\pc}{{\rm pc}}

\newcommand{\rh}{R_{\rm h}}
\newcommand{\rhdot}{\dot{R}_{\rm h}}

\newcommand{\rhoh}{\rho_{\rm h}}
\newcommand{\rhn}{R_{\rm h0}}

\newcommand{\rv}{R_{\rm v}}
\newcommand{\rvn}{R_{\rm v0}}
\newcommand{\tcc}{T_{\rm cc}}

\newcommand{\tdyn}{T_{\rm dyn}}

\newcommand{\trh}{T_{\rm rh}}
\newcommand{\trhn}{T_{\rm rh0}}
\newcommand{\tms}{t_{\rm ms}}

\newcommand{\ts}{T_*}

\documentclass[useAMS,usenatbib, times]{mn2e}
\usepackage{amssymb,latexsym,graphicx,natbib,eufrak,times}

\title[On the mass-radius relation of hot stellar systems]
  {On the mass-radius relation of hot stellar systems}
\author[Gieles, Baumgardt, Heggie \& Lamers]
  {Mark~Gieles$^{1,2,3}$, Holger~Baumgardt$^{4,5}$, Douglas~C. Heggie$^2$ and Henny~J.G.L.M.~Lamers$^6$\\
  $^1$ Institute of Astronomy, University of Cambridge, Madingley Road, Cambridge, CB3 0HA, UK\\
  $^2$ School of Mathematics and Maxwell Institute for Mathematical Sciences, University of Edinburgh, King's Buildings, Edinburgh, EH9 3JZ, UK\\
  $^3$ European Southern Observatory, Casilla 19001, Santiago 19, Chile \\
  $^4$ Argelander Astronomical Institute, University of Bonn, Bonn, Germany\\ 
  $^5$ School of Mathematics and Physics, The University of Queensland, Brisbane, QLD 4072, Australia\\
  $^6$ Astronomical Institute, Utrecht University, 
  Princetonplein 5, NL-3584CC Utrecht, the Netherlands
}
\date{Accepted 2010 July 14.  Received 2010 July 13; in original form 2010 June 24}

\def\LaTeX{L\kern-.36em\raise.3ex\hbox{a}\kern-.15em
    T\kern-.1667em\lower.7ex\hbox{E}\kern-.125emX}

\begin{document}         
\maketitle
\begin{abstract}
Most globular clusters have half-mass radii of a few pc with no
apparent correlation with their masses. This is different from
elliptical galaxies, for which the Faber-Jackson relation suggests a
strong positive correlation between mass and radius.  Objects that are
somewhat in between globular clusters and low-mass galaxies, such as
ultra-compact dwarf galaxies, have a mass-radius relation consistent
with the extension of the relation for bright ellipticals.  Here we
show that at an age of $10\,\gyr$ a break in the mass-radius relation
at $\sim10^6\,\msun$ is established because objects below this mass,
i.e. globular clusters, have undergone expansion driven by stellar
evolution and hard binaries.  From numerical simulations we find that
the combined energy production of these two effects in the core comes
into balance with the flux of energy that is conducted across the
half-mass radius by relaxation.  An important property of this
`balanced' evolution is that the cluster half-mass radius is
independent of its initial value and is a function of the number of
bound stars and the age only.  It is therefore not possible to infer
the initial mass-radius relation of globular clusters and we can only
conclude that the present day properties are consistent with the
hypothesis that all hot stellar systems formed with the same
mass-radius relation and that globular clusters have moved away from
this relation because of a Hubble time of stellar and dynamical
evolution.
\end{abstract}

\section{Introduction}
The half-mass radius of old globular clusters in the Milky Way depends
 only weakly on mass \citep[e.g.][]{1991ApJ...375..594V}. If anything,
 a negative correlation between radius and mass is found for the
 clusters in the outer halo \citep{2004MNRAS.354..713V}.  Because this
 is also found for extra-galactic globular
 clusters \citep{2005ApJ...634.1002J, 2007AJ....133.2764B,
 2009MNRAS.392..879G} the mass-radius relation, or lack thereof, is an
 important aspect of the fundamental plane relations of globular
 clusters \citep{1995ApJ...438L..29D,2000ApJ...539..618M}.

Objects more massive than typical globular clusters, such as the
 recently discovered ultra-compact dwarf
 galaxies \citep[UCDs,][]{1999A&AS..134...75H, 2003Natur.423..519D},
 but also the most massive globular clusters, do exhibit a positive
 correlation between radius and
 mass \citep{2005ApJ...627..203H,2007A&A...469..147R,2008A&A...487..921M}.
 Interestingly, the position of systems more massive than
 $\sim10^6\,\msun$ in the mass-radius diagram coincides with the
 extension to low masses of the \citet{1976ApJ...204..668F} relation
 for bright elliptical galaxies \citep{2005ApJ...627..203H}. The
 mass-radius relation of stellar systems more massive than
 $\sim10^6\,\msun$ has been explained by the details of their
 formation \citep{2009ApJ...691..946M}, where this was considered a
 deviation from the near constant radius of less massive systems.  In
 this study we test the hypothesis that all hot stellar systems
 (globular clusters, UCDs and elliptical galaxies) had the same
 mass-radius relation initially and that the globular clusters
 ($\lesssim10^6\,\msun$) are the deviators because they have
 moved away from this relation because of dynamical evolution.

Intuitively we can expect that low mass stellar
systems are dynamically more evolved than massive systems because of
their shorter relaxation time-scale. This evolutionary time-scale is
often expressed in terms of the half-mass properties of the
system \citep{1987degc.book.....S}

\begin{equation}
\trh=0.138\frac{N^{1/2}\rh^{3/2}}{G^{1/2}\bar{m}^{1/2}\ln\Lambda},
\label{eq:trh}
\end{equation}
where $N$ is the number of stars, $\rh$ is the half-mass radius, $G$
is the gravitational constant, $\bar{m}$ is the mean stellar mass and
$\Lambda$ is the argument of the Coulomb logarithm and equals
$0.02N\lesssim\Lambda\lesssim0.11N$ depending on the stellar mass
function in the cluster \citep{1994MNRAS.268..257G}.  If we take the
initial mass-radius relation to be of the form $\rhn\propto
M_0^\lambda$, then $\trhn$ is an increasing function of $M_0$ for all
$\lambda>-1/3$. Although the value of $\lambda$ is poorly constrained
from observations, it is unlikely to be negative and we can,
therefore, safely say that low mass stellar systems have shorter
relaxation times than massive systems immediately after formation.

Here we consider the expansion of star clusters driven by mass loss
due to stellar evolution and hard binaries and we present a
description for the radius evolution including both effects, based on
results of $N$-body simulations
(\S~\ref{sec:expansion}). In \S~\ref{sec:application} we show that at
an age of $10\,\gyr$ a Faber-Jackson type initial mass-radius relation
has been erased because of the expansion of stellar sytems with
$M\lesssim10^6\,\msun$. A summary and discussion is presented
in \S~\ref{sec:summary}.

\section{Expansion of stellar systems}
\label{sec:expansion}
We want to understand the evolution of the radius of a stellar system
with a realistic stellar mass function in which the stars evolve and
lose mass in time. This evolution is distinct from the well studied
and well understood behaviour of an equal-mass
cluster \citep[e.g.][]{1965AnAp...28...62H,
1984ApJ...280..298G}. Because we are mainly interested in the
expansion we ignore the effect of a tidal cut-off.  As we will show
in \S~\ref{sec:application}, the results explain the mass-radius
relation of objects with $M\gtrsim10^5\,\msun$, suggesting that tides
are not very important in shaping the mass-radius relation of these
objects. We first consider various stellar mass functions, ignoring
the effect of stellar evolution (\S~\ref{ssec:mf}), and then add the
effect of stellar evolution in \S~\ref{ssec:mf_sev}.

\subsection{Expansion driven by hard binaries}
\label{ssec:mf}
The evolution of equal-mass clusters has been studied in quite some
detail \citep[e.g.][]{1994MNRAS.268..257G,2002MNRAS.336.1069B}. To
first order their entire evolution follows from the fact that
gravitational systems have negative total energy, which causes them to
always evolve away from thermal equilibrium. In the early evolution
this results in a contraction of the core and this inevitably leads to
the gravothermal catastrophe, or core
collapse \citep[][]{1980MNRAS.191..483L}.  For an
equal-mass \citet{1911MNRAS..71..460P} model the time of core-collapse
is at
$\tcc\approx17\,\trhn$ \citep[e.g.][]{1987degc.book.....S}. After core
collapse the evolution is driven by binaries in the core that release
energy to the rest of the cluster when they form and harden in 3-body
interactions. This energy is conducted outwards by 2-body relaxation
and in the absence of a tidal field this results in an expansion of
the cluster as a whole, because escape of stars is inefficient.  This
increase of $\rh$ happens on a relaxation time-scale such that we can
say $\rhdot=\zeta\rh/\trh$.  If we integrate this relation from $\tcc$
to $T$, taking into account the $\rh$ dependence in $\trh$
(equation~\ref{eq:trh}) we find
\begin{eqnarray}
\rh&=&\rhn\left(1+\frac{\chi [T-\tcc]}{\trhn}\right)^{2/3}, \\
     &\approx&\rhn\left(\frac{\chi T}{\trhn}\right)^{2/3},
\label{eq:rh}
\end{eqnarray}
where $\chi$ is a constant that relates to $\zeta$ as
 $\chi\equiv(3/2)\zeta$. In the last step we have used
 $\tcc\approx\trhn/\chi$ as the integration boundary, which is not
 strictly true.  For an equal-mass cluster
 $\chi\approx0.14$ \citep{1965AnAp...28...62H, 2003gmbp.book.....H}
 and, therefore, $1/\chi\approx7.2$, whereas for the Plummer model we
 have $\tcc/\trhn\approx17$ \citep[see
 also][]{1994MNRAS.270..298G}. But equation~(\ref{eq:rh}) describes
 the asymptotic behaviour of $\rh$ for $T>>\tcc$ and is therefore a
 useful approximation. It also follows from equation~(\ref{eq:rh})
 that after $\tcc$ the evolution of $\rh$ is independent of $\rhn$: if
 we assume that $N$ and $\mmean$ do not change in time then we can say
 $\trhn=\trh\rhn^{3/2}/\rh^{3/2}$ (equation~\ref{eq:trh}) and
 equation~(\ref{eq:rh}) is equivalent to $\trh=\chi T$.  This means
 that after some time clusters evolve towards a mass-radius relation
 of the form $\rh\propto T^{2/3} M^{-1/3}$, independent of the initial
 mass-radius relation. This is an important result and
 in \S~\ref{sec:application} we will show that it applies to globular
 clusters.

The presence of a mass function speeds up the dynamical
evolution \citep{1985ApJ...292..339I}, in the sense that core collapse
happens earlier \citep{2004ApJ...604..632G} and the escape rate of
clusters in a tidal field is higher \citep{1995ApJ...443..109L}.  Here
we establish by means of direct $N$-body simulations how the rate of
expansion, i.e. the value of $\chi$, depends on the mass function of
the stars.  We consider a \citet{2001MNRAS.322..231K} stellar mass
function and vary $\mu\equiv\mmax/\mmin$, where $\mmax$ and
$\mmin=0.1\,\msun$ are the maximum and minimum stellar mass,
respectively\footnote{We use the ratio $\mmax/\mmin$ because it is
easy to relate to real clusters. \citet{2004ApJ...604..632G} show that
the relevant parameter is $\mmax/\mmean$ which captures variations in
$\mmax$ and the slope of the mass function.}. We consider values from
$\mu=1$ (=\,equal-mass) to $10^3$ (=\,full mass function) in steps of
a factor of 10.  These values cover the relevant values of $\mu$ for
real clusters.  We model clusters with $N=4096, 8192, 16384$ and
$32768$ particles and multiple runs are done for clusters with low $N$
and/or high $\mu$ to average out statistical fluctuations due to the
low number of (massive) stars. The number of simulations was chosen to be
$\max[1,(32768/N)/(5-\log\mu)]$.  The initial density profile of all
clusters is described by Plummer models in virial equilibrium and
during the simulation stars are taken out of the simulation when they
reach 20$\,\rv$, where $\rv$ is the virial radius.  The models are all
scaled to the usual $N$-body units \citep[$G=\rvn=-4\,E_0=1$, where
$E_0$ is the total initial energy, ][]{1986LNP...267..233H} and we use
the \texttt{kira} integrator which is part of the \texttt{Starlab}
software \citep{2001MNRAS.321..199P} to numerically solve the $N$-body
problem in time.  At each time the values for $\rh$, $\mmean$ and $N$
are recorded and $\trh$ is calculated using equation~(\ref{eq:trh}).
For $\mu=1$ we use $\gamma=0.11$ while for $\mu>1$ we use
$\gamma=0.02$ as recommended by \citet{1994MNRAS.268..257G}.

\begin{figure}
 \includegraphics[width=8.cm]{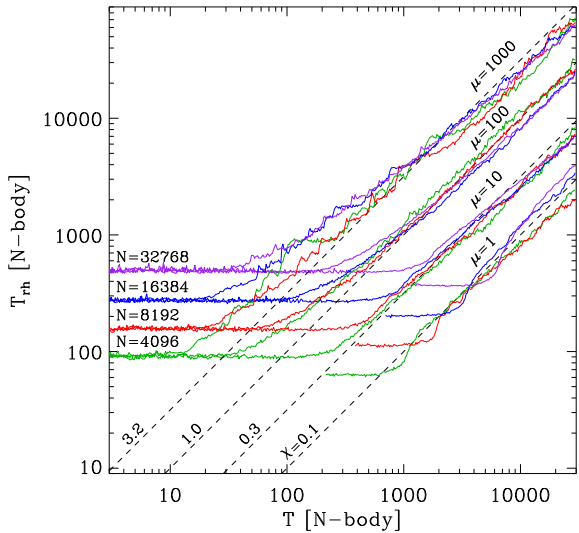}     
 \caption{Evolution of the half-mass relaxation time, $\trh$, for clusters with
     different $N$ and different $\mu$. The $N$-body unit of time,
     $\tdyn$, can be related to physical units through
     $\tdyn=(GM/\rv^3)^{-1/2}$. A Kroupa~(2001) mass function is used
     for the stars in the range $0.1\le m/\msun \le0.1\mu$. Clusters
     of different $N$ and the same $\mu$ evolve to the same
     $\trh\approx\chi T$ after core-collapse
     (equation~\ref{eq:rh}). This asymptotic behaviour is roughly
     matched by the relation $\chi\approx 0.1\mu^{1/2}$, shown as
     dashed lines. The $\trh$ values of the equal mass clusters are
     calculated using a slightly different argument in the Coulomb
     logarithm ($\Lambda=0.11N$, equation~\ref{eq:trh}) as compared to
     the multi-mass clusters ($\Lambda=0.02N$). For clarity the
     $\mu=1$ curves are only plotted for
     $T\gtrsim3\trhn$.}  \label{fig:expansion_nosev}
\end{figure}

In Fig.~\ref{fig:expansion_nosev} we show the (average) resulting
evolution of $\trh$ for all 16 different initial conditions, specified
by the number of stars $N$ and the width of the stellar mass function,
$\mu$. The increase of $\trh$ after $T\approx\trhn/\chi$ is dominated
by expansion, because $\mmean$ remains constant (no stellar evolution)
and the number of bound stars does not change much. The dashed lines
indicate different values of $\chi$ and it can be seen that the
dependence of $\chi$ on the mass function can to first order be
approximated by $\chi\approx0.1\mu^{1/2}$ (or
$\chi\approx0.1(\mmax/\mmean)^{0.7}$). This scaling roughly
recovers \henon's result for equal-mass models. In summary, we see
from Fig.~\ref{fig:expansion_nosev} that $T/\trh$ increases until
$T\approx\trhn/\chi$, after which $T/\trh\approx\,$constant.

In \S~\ref{ssec:mf_sev} we repeat the simulations with $\mu=10^3$ and
turn stellar evolution on such that $\mu$ naturally decreases from
$10^3$ at $T=0$ to $\mu\approx10$ at $T\approx10\,\gyr$ during the
simulation because of stellar evolution.

\subsection{The combined effect of  stellar evolution and binaries}
\label{ssec:mf_sev}
The time-scales of stellar evolution are set by the stellar interiors
and are independent of the relaxation time-scale of the cluster
wherein the stars evolve.  We thus expect the details of the evolution
to depend on a combination of the stellar evolution time-scale and the
relaxation time of the cluster.  Here we show that the resulting
expansion still depends in a simple way on the dynamical properties of
the cluster as a whole.
 
We want to consider a large range of $\trhn$ with our simulations to
cover a parameter space that is relevant for real globular clusters.
Because computing times limit us to $N\lesssim10^5$ with direct
$N$-body simulations, we vary both $N$ and the initial half-mass
density, $\rhoh\equiv 3M/(8\pi\rh^3)$. We consider 15 different values
of $\trhn$ ranging from $\trhn\approx1\,\myr$
$([N,\log\rhoh]=[8192,6])$ to $\trhn\approx4\,\gyr$
$([N,\log\rhoh]=[131072,1])$, with $\rhoh$ in $\msunpc$. Here $\trhn$
is increased by increasing $N$ by factors of $2$ and by decreasing
$\rhoh$ by factors of $10$. We again use the \texttt{kira} integrator
and the stellar evolution package \texttt{SeBa} for solar
metallicity \citep{2001MNRAS.321..199P}.  We use a Kroupa~(2001)
initial mass function between $0.1\,\msun$ and $100\,\msun$, which has
$\mmean\approx0.64\,\msun$. The retention fraction of black holes and
neutron stars was set to zero.

\begin{figure}
 \includegraphics[width=8.cm]{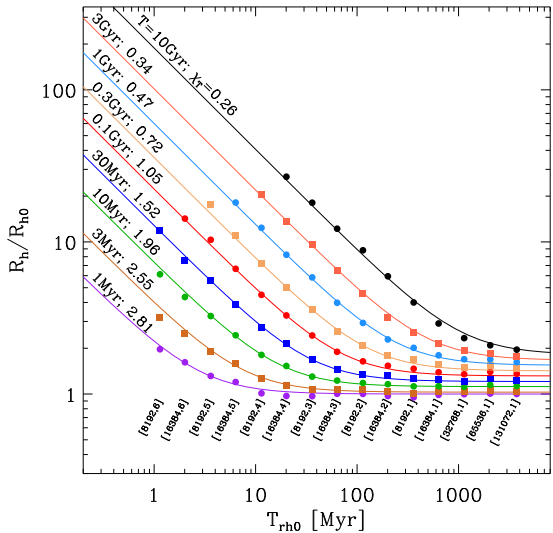} 
 \caption{Expansion from the $N$-body runs including the effect of
   stellar evolution together with the functional fits
   (equation~\ref{eq:rhtot}, full lines)} \label{fig:exp}
\end{figure}

In Fig.~\ref{fig:exp} we show the resulting expansion in the form of
$\rh/\rhn$ as a function of $\trhn$ at different ages. The asymptotic
behaviour of these runs can easily be understood by considering the
extremes. Clusters that are dynamically young (low $T/\trhn$) expand
adiabatically in order to retain virial equilibrium after stellar mass
loss.  The continuous loss of mass from a \citet{2001MNRAS.322..231K}
mass function together with the stellar evolution prescription
of \texttt{Starlab} \citep[Appendix~B2 of][]{2001MNRAS.321..199P}
leads to a reduction of the total cluster mass
\begin{equation}
M\approx M_0\left(\frac{T}{\ts}\right)^{-\delta}, \,\,\,T\ge\ts,\,\,\, \delta\approx0.07, \,\,\,\ts\approx2\,\myr.
\label{eq:mt}
\end{equation}
In this regime the radius thus evolves as \citep[e.g.][]{1980ApJ...235..986H}
\begin{equation}
\rh\approx \rhn\left(\frac{T}{\ts}\right)^{\delta}.
\label{eq:rt}
\end{equation}
This adiabatic expansion is slow in time and gives a maximum increase
of $\rh/\rhn\approx2$ after a Hubble time.  At the other extreme we
have clusters that are dynamically old (high $T/\trhn$) and they
expand quickly in a way that is comparable to what we have seen
in \S~\ref{ssec:mf}.  We propose a function that stitches together
these two extremes in an attempt to match $\rh/\rhn$ for all values of
$T/\trhn$

\begin{equation}
\rh=\rhn\left(\left[\frac{T}{\ts}\right]^{2\delta}  + \left[\frac{\chifit T}{\trhn}\right]^{4/3}\right)^{1/2},  T\ge\ts.
\label{eq:rhtot}
\end{equation}
Here $\chifit$ is a parameter comparable to $\chi$
of \S~\ref{ssec:mf}, but now time-dependent due to the variation of
the mass function, and its value at an age $T$ is found from a fit of
equation~(\ref{eq:rhtot}) to the results of the $N$-body runs. In
Fig.~\ref{fig:exp} we show the fit results as full lines and the
resulting values of $\chifit$ are indicated. It shows that
equation~(\ref{eq:rhtot}) provides a good description of the evolution
of $\rh/\rhn$. The relation between $\chifit$ and $T$ is well
approximated by a simple power-law function

\begin{equation}
\chifit \approx 3\left(\frac{T}{\ts}\right)^{-0.3}, \,\,\,\ts\le T\lesssim20\,\gyr.
\label{eq:chifit}
\end{equation}
If we now define $\ts\equiv\min([2\,\myr,T])$ we have a continuous
function for $\rh(\trhn,T)$, or $\rh(M_0,\rhn,T)$ for all $T$.  For
high $T/\trhn$ we find from equation~(\ref{eq:rhtot}) that
$\trh=\left(\mmeann/\mmean\right)^{1/2}\chifit T\propto T^{0.74}$
(equations~\ref{eq:mt}\,\&\,\ref{eq:chifit}).  We indicate below how
the small deviation from a linear scaling with $T$ (as found
in \S~\ref{ssec:mf}) can be interpreted in terms of the evolution of
the mass function. In \S~\ref{ssec:mf} we found
$\trh\propto\mu^{1/2}T$. If we approximate the main sequence life-time
of stars by $\tms\propto m^{-2.5}$ \citep{1993A&AS..100..647B} and
thus $\mu\propto T^{-1/2.5}$, then from the changing mass function we
expect $\trh\propto T^{0.8}$, very close to what we find from the
numerical simulations. We conclude that the evolution of the cluster
is `balanced' when the second term on the right-hand side of
equation~(\ref{eq:rhtot}) dominates (high $T/\trhn$), in the sense
that the energy flux at the half-mass boundary that drives the
expansion is provided by the production of energy in the core by
binaries and stellar evolution combined. Stellar evolution in fact
slows down the expansion rate, because $\rhdot$ is determined by the
instantaneous width of the stellar mass function, resulting in a
smaller $\rhdot$ at old ages than if $\mu$ had stayed constant at
$\mu=10^3$ (\S~\ref{ssec:mf}).
When the first term on the right-hand
side of equation~(\ref{eq:rhtot}) dominates (low $T/\trhn$)
 the evolution is unbalanced and we have
the usual adiabatic expansion .
 
The fact that the interplay between dynamical evolution and stellar
evolution is in fact quite simple can be understood from the energy
budget.  The total energy of a stellar systems depends on $M$ and
$\rh$ as $E\propto -M^2/\rh$.  Together with
equations~(\ref{eq:mt})\,\&\,(\ref{eq:rt}) we find that $E$ evolves in
time as $E\propto -(T/\ts)^{-3\delta}$ because of mass-loss and
(adiabatic) expansion. The rate of energy change as a result of
stellar evolution is then $\edotsev\propto |E|/T$, where the constant
of proportionality depends on the degree of mass segregation: it will
be higher when stars lose mass from the centre and/or when the density
profile is centrally concentrated.  The rate of energy increase due to
binaries and relaxation is similar. This is because $\edot\propto
|E|/\trh$ and $\trh\propto T$ (\S~\ref{ssec:mf}).  After some
dynamical relaxation mass-loss by stellar evolution will be
predominantly from the core because of mass segregation and this will
boost $\edotsev$.  We tentatively pose the idea that $\edotsev$ acts
as a central energy source that is subject to a feedback mechanism
comparable to what happens with binaries: if $\edotsev$ is too high,
the core expands and the central potential decreases and $\edotsev$
drops. If $\edotsev$ is too low, the core contracts, thereby
increasing the depth of the central potential and increasing
$\edotsev$. This fits in the view of \citet{1975IAUS...69..133H} that
``the rate of flow of energy is controlled by the system as a whole,
not by the singularity''.  One of the consequences is that there is no
sharp transition between a stellar evolution dominated phase and a
relaxation dominated phase.

\section{Application to old stellar systems}
\label{sec:application}
With the expression for the evolution of the radius as a function of
$\trhn$ at hand we can easily calculate the evolution of $\rh$ for any
initial mass-radius relation. We apply our result to the mass-radius
relation of old and hot stellar systems in the mass range
$~\sim10^4-10^8\,\msun$.

The original Faber-Jackson relation relates the central velocity
dispersion of (bright) elliptical galaxies to their total
luminosity. \citet{2005ApJ...627..203H} have converted this result
into relations between $M$, $\rh$ and surface density. The resulting
mass-radius relation (their equation~15) with an additional $\log 4/3$
to correct for projection is
$\log(\rh/\pc)=-3.142+0.615\log(M/\msun)$. They show that this
relation matches the objects with $M\gtrsim 10^6\,\msun$ (UCDs,
massive globular clusters and their dwarf-globular transition objects,
DGTOs) in the mass-radius diagram. Because this concerns
collision-less systems, we can safely assume that 2-body relaxation
has not affected this relation and it should, therefore, reflect the
initial relation. To get an expression for the initial mass-radius
relation we only need to correct for mass-loss by stellar evolution
and the subsequent adiabatic expansion. For $T=10\,\gyr$, we find
$M/M_0=\rhn/\rh\approx0.55$ (equations~\ref{eq:mt}\,\&\,\ref{eq:rt})
and thus

\begin{equation}
\log\left(\frac{\rhn}{\pc}\right)=-3.560+0.615\log\left(\frac{M_0}{\msun}\right).
\end{equation}
In Fig.~\ref{fig:mass_radius_data} we show how this initial
mass-radius relation evolves using our result from
equation~(\ref{eq:rhtot}) together with data points that cover the
mass regime we are interested in. For high $T/\trhn$ the radius is set
by $M_0$, independent of $\rhn$, while for low $T/\trhn$ we are seeing
roughly the initial mass-radius relation. By construction the
right-hand side of the 10\,\gyr\ line coincides with the
representation of the Faber-Jackson relation
of \citet{2005ApJ...627..203H}.  From solving $\dr\rh/\dr M=0$ in
equation~(\ref{eq:rhtot}) we find that at an age of $10\,\gyr$ the
break between the two regimes occurs at
$M_0\approx1.1\times10^6\,\msun$ and at that age systems with this
mass have $\trh/T\approx0.8$.
\citet{2008A&A...487..921M} noticed already that the break occurs at systems with $\trh$ roughly equal to a Hubble time. In this
paper we give a quantitative explanation for it.

\section{Summary and discussion}
\label{sec:summary}
In this study we provide the arguments that explain why there is a
break in the mass-radius relation of hot stellar systems at
$\sim10^6\,\msun$.  We show that the mass-radius relation of the
massive systems ($\gtrsim10^6\,\msun$) is only slightly affected by
stellar evolution and represents, therefore, approximately the initial
mass-radius relation.  The origin of this relation needs to be
searched for in the details of their formation and is not discussed
here \citep[see e.g.][]{2009ApJ...691..946M}.  Combining scaling
relations for the (adiabatic) expansion of clusters because of stellar
evolution with relations for expansion due to 2-body relaxation we
present a simple formula for the radius evolution as a function of
initial mass, radius and time. Applying this result to a Faber-Jackson
type initial mass-radius relation \citep[the representation in units
of mass and radius are taken from][]{2005ApJ...627..203H} we show that
at an age of $10\,\gyr$ a break occurs at $\sim10^6\,\msun$.  This
break can be thought of as the boundary between collisional systems
($\trh\lesssim$\,age) and collision-less systems ($\trh\gtrsim$\,age).

\begin{figure}
 \includegraphics[width=8.cm]{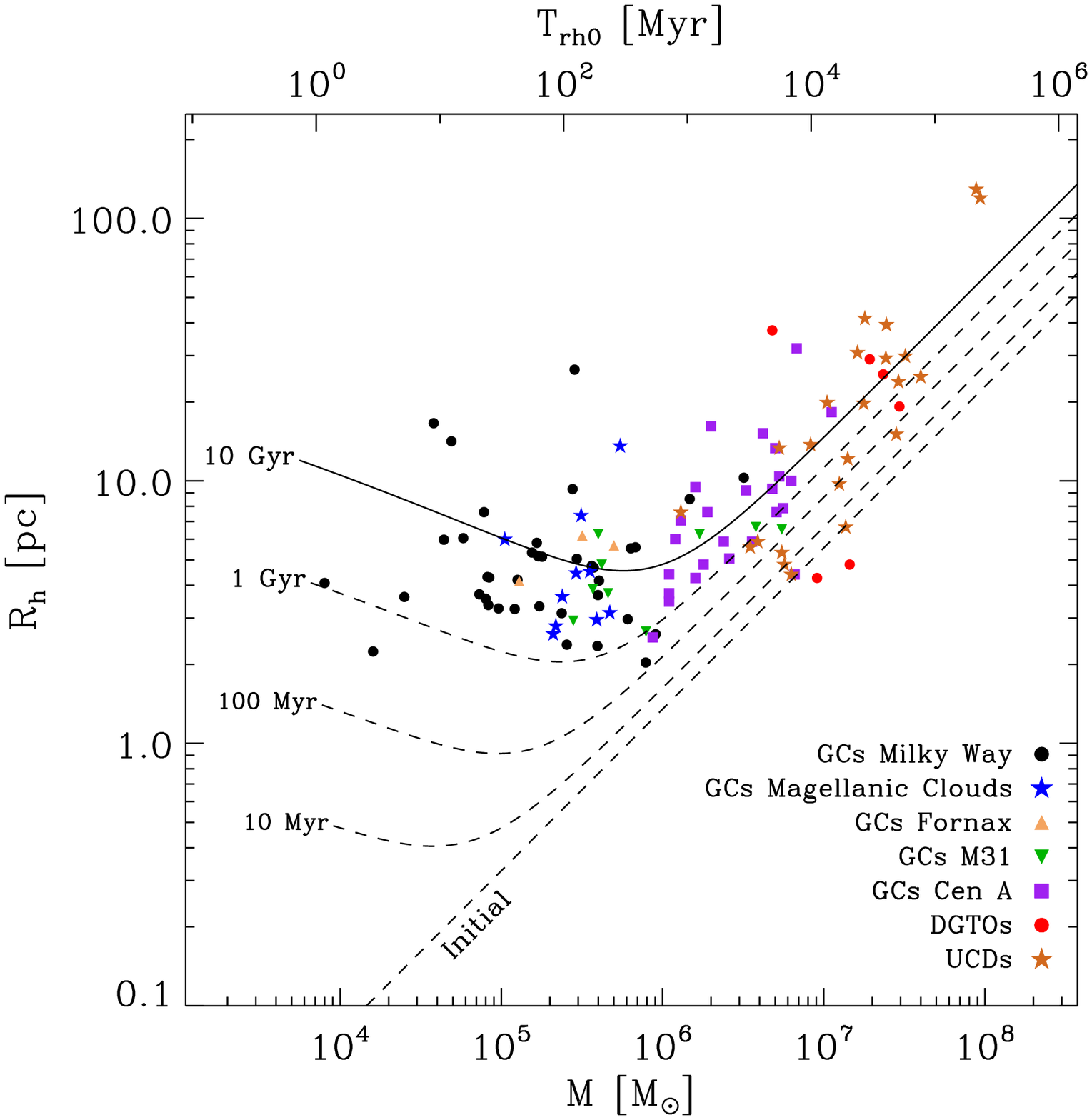} 
 \caption{Mass-radius values for hot stellar systems.  The values for
     globular clusters in the Milky Way, the Magellanic Clouds and
     Fornax are taken from \citet{2005ApJS..161..304M}. The clusters
     in M31 are from \citet{1997A&A...321..379D}. The values for
     globular clusters in NGC~5128 (Cen~A), UCDs and DGTOs are from
     the compilation presented in \citet{2008A&A...487..921M}. The
     lines show the evolution of the mass-radius relation using the
     Faber-Jackson relation, corrected for stellar evolution, as
     initial conditions. The break at $\sim10^6\,\msun$ at
     $T\approx10\,\gyr$ is because lower mass objects have expanded.
     } \label{fig:mass_radius_data}
\end{figure}

For young massive clusters there is also no obvious correlation
between radius and mass/luminosity
\citep[][]{1999AJ....118..752Z,2004A&A...416..537L,2007A&A...469..925S,2010arXiv1002.1961P}. 
From Fig.~\ref{fig:mass_radius_data} it can be seen that for clusters
with an age of $\sim10-100\,\myr$ there has already been significant
expansion of clusters with masses $\lesssim10^5\,\msun$. Although this
break mass depends on the initial mass-radius relation, it at least
qualitatively shows that at young ages most clusters\footnote{The
luminosity function of young clusters is a power-law distribution with
index $\sim-2$ such that a typical young cluster population only has a
small fraction of its clusters in the massive ($\gtrsim10^5\,\msun$)
tail.} are affected by the expansion we consider here. It is
worthwhile to compare the theory to the parameters of young, well
resolved star clusters \citep[e.g.][]{2003MNRAS.338...85M}.  We
emphasise that the balanced evolution provides a lower limit to
cluster radii. If clusters form above the relation marked initial in
Fig.~\ref{fig:mass_radius_data} then they expand only slightly because
of stellar evolution at young ages until $T/\trhn$ is high enough for
the balanced evolution/expansion to start. The mass-radius relation of
young clusters is important in the evolution of cluster populations.
This is because in the early evolution clusters suffer from encounters
with the molecular gas clouds from which they form. The time-scale of
disruption due to such encounters scales with the density of the
cluster \citep{1958ApJ...127...17S}. If all cluster have the same
density, their disruption time-scale is independent of their mass. For
a constant radius the time-scale of disruption becomes strongly
mass-dependent because then $\rhoh\propto M$ and for a constant $\trh$
we have $\rhoh\propto M^2$. The mass-radius relation, therefore,
determines the properties of the clusters that survive continuous
encounters with massive clouds \citep{2006MNRAS.371..793G,
2010ApJ...712L.184E}.
 
We have ignored the tidal limitation due to the host galaxy. Once the
density of a cluster drops below a critical value, depending on the
tidal field strength, our result will overestimate the radius of such
clusters because the presence of a tidal limitation will prevent
further growth.  The good agreement between the simple model presented
here and the data points suggest that at least to first order the
positions in the mass-radius diagram of objects with
$M\gtrsim\,$few$\times10^4\,\msun$ is not much affected by
tides. Including a tidal field would bend down the curves at low
masses. The transition from expansion dominated evolution to
Roche-lobe filling evolution is considered in more detail in a
follow-up study \citep*{G10b}.

\vspace{-0.5cm}
\section*{Acknowledgement}
MG thanks the Royal Society for financial support.  The simulations
were done on the GRAPE-6~BLX64 boards of the European Southern
Observatory in Garching.  This research was supported by the DFG
cluster of excellence Origin and Structure of the Universe
(www.universe-cluster.de).  HB and HJGLM thank ESO for a Visiting
Scientist Fellowship in Santiago in 2009 where this project was
started.

\vspace{-0.5cm}
\bibliographystyle{mn2e}

\end{document}